\documentclass[aps,prd,groupedaddress,nofootinbib,amssymb,balancelastpage]{revtex4}

\usepackage[dvipdfmx]{graphicx}
\usepackage{graphicx,bm,color}
\usepackage{amsmath}
\usepackage{amssymb}
\usepackage{amsfonts}
\usepackage{cases}
\usepackage{cancel}
\usepackage{hyperref}
\usepackage{ulem}

\newcommand{\cred}[1]{{\color{black}#1}}

\begin{document}

\title{
Unified Interpretation of Scalegenesis in Conformally Extended Standard Models:\\ 
A Dynamical Origin of Higgs Portal}

\author{Hiroyuki Ishida}
\email{ishidah@post.kek.jp}
\affiliation{Theory Center, IPNS, KEK, Tsukuba, Ibaraki 305-0801, Japan}
\author{Shinya Matsuzaki}
\email{synya@jlu.edu.cn}
\affiliation{Center for Theoretical Physics and College of Physics, Jilin University, Changchun, 130012, China}
\author{Ruiwen Ouyang}
\email{ruiwen.ouyang@kbfi.ee}
\affiliation{Laboratory of High Energy and Computational Physics, National Institute of Chemical Physics and Biophysics, R$\ddot{a}$vala pst. 10, 10143 Tallinn, Estonia}

\begin{abstract}
We present a universal interpretation for a class of conformal extended standard models 
including Higgs portal interactions realized in low-energy effective theories. 
The scale generation mechanism in this class (scalegenesis) 
arises along the (nearly) conformal/flat direction for the scale symmetry breaking,  
where the electroweak-symmetry breaking structure is achieved in a similar way to the standard model's. 
A dynamical origin for the Higgs portal coupling can provide 
the discriminator for the low-energy ``universality class'', 
to be probed in forthcoming collider experiments. 
\end{abstract}

\maketitle


\section{Introduction} 

By the Higgs discovery~\cite{Aad:2012tfa,Chatrchyan:2012xdj}, 
the standard model (SM) of particle physics has been completed. 
It is however still unsatisfactory that  
the detailed dynamics of the electroweak symmetry breaking (EWSB) is not unraveled yet:  
in the SM,  
the sign of the Higgs mass is necessarily 
assumed to be negative to realize EWSB, 
so in that sense the SM just gives an incomplete answer 
to the origin of the EWSB as well as the origin of mass. 
This issue would be related to the gauge hierarchy problem or fine tuning problem involving 
the physics bridging over the EW and Planck scales 
through the unique dimensionful parameter. 
Motivated by this longstanding problem, 
people have so far extensively been working on a new dynamics and/or mechanism 
which could be dormant behind the Higgs sector.

It is a scale symmetry  
that could be one of the clues to 
access this issue and has currently been major in the ballpark of the Higgs physics,  
as a possible solution to the gauge hierarchy problem, 
a la Bardeen~\cite{Bardeen:1995kv}: 
Quadratic divergent corrections to the Higgs mass term, the critical 
part for the hierarchy problem, 
are assumed not to give a physical scaling, hence should be 
removed, so that 
the Higgs mass term only gets logarithmic corrections proportional to 
the bare Higgs mass or SM particle masses coupled to the Higgs. 
In that case, no gigantic cancellation or instability against the radiative 
power corrections associated with the Planck scale, 
is required for the Higgs mass term 
-- no gauge hierarchy problem is present. 
If the Higgs mass parameter can be turned off 
at some scale in the renormalization evolution, say, at 
the Planck scale, thus the Higgs mass will not develop 
up to the logarithmic corrections. 
This can be done by assuming realization of the scale symmetry at 
the Planck scale, and 
the physical Higgs mass then may arise by feeding only the logarithmic 
corrections as the quantum scale anomaly effect.

Nature might have in fact supported the presence of an approximate scale 
(or conformal) invariance, 
and an {\it orientation} nearly along a conformal theory:  
the observed SM-like Higgs is thought to be lying along a nearly conformal direction in the EW broken phase with the vacuum expectation value 
(VEV) $v \simeq 246$ GeV, and acquires the mass 
due to the small quartic coupling breaking the scale symmetry 
(at the tree-level). 
$\lambda_H = (m_h^2/2v^2) \simeq 1/8 \ll 1$.  
Thus a {\it flat (conformal) direction} can be seen in the 
SM by taking the limit $\lambda_H 
\to 0$, where the Higgs potential in the EW broken phase gets completely flat.

Generically, flat (conformal) directions are observed as 
stationary hyper-surfaces spanned by aligned scalar VEVs: $v_i = n_i v$, where $i$ runs the number of scalars and $v$ is the average magnitude of the VEVs spontaneously breaking the scale symmetry~\cite{Gildener:1976ih}. 
Along the flat (conformal) direction, one finds 
the flat curvature, hence expects the presence of a massless scalar 
associated with the scale (conformal) symmetry broken by the 
VEVs, i.e., ``dilaton''. 
The conformal limit in the SM ($\lambda_H \to 0$)  
can be understood as the simplest case for a generic flat direction 
argument~\cite{Kannike:2019upf}, 
where the zero determinant of the quartic coupling matrix, reflected as the constraint on the $n_i$ vector, 
is given as a necessary condition to have a flat direction, 
which is trivially realized by the $\lambda_H =0$ in the case of the SM.   
In that sense, the SM can indeed be dubbed as a nearly (classically) conformal theory (the nearly conformal SM), and the 125 GeV Higgs can be 
regarded as a light ``pseudo-dilaton'' associated with the approximate 
conformal direction (with the small curvature).

This observation is still perturbatively operative even including quantum corrections, which will however directly be linked with the instability of the EW vacuum in the SM 
when 
the $\lambda_H$ gets smaller, even approaches or goes negative because of the sizable top loop correction~\cite{Cabibbo:1979ay,Hung:1979dn,Lindner:1988ww,Arnold:1989cb,Sher:1988mj,Sher:1993mf,Schrempp:1996fb,Arnold:1991cv,Casas:1994qy,Casas:1996aq,Hambye:1996wb,Isidori:2001bm,Einhorn:2007rv,Ellis:2009tp,Gerhold:2009ub,Holthausen:2011aa,EliasMiro:2011aa,Bezrukov:2012sa,Degrassi:2012ry,Alekhin:2012py,Masina:2012tz,Hamada:2012bp,Jegerlehner:2013cta,Buttazzo:2013uya,Branchina:2013jra,Kobakhidze:2014xda,Spencer-Smith:2014woa,Branchina:2014usa,Branchina:2014rva,Bednyakov:2015sca}.   
Thus, the nearly conformal SM has to be cured by some new dynamics, 
which stays along a conformal direction including the SM as 
the low-energy description, keeping the approximate 
scale invariance (without power corrections to 
Higgs mass term) at the quantum level up until the Planck scale.

The realization of scale invariance at the quantum level \cred{(quantum scale-invariance)}
has been developed from Bardeen's initial proposal as described above, 
which currently is not just an
ad hoc assumption, rather, involves two nontrivial dynamical
issues: One is to dynamically 
achieve the initial renormalization condition at the Planck scale $(\Lambda_{\rm pl})$ 
as the Higgs mass parameter $m_H(\Lambda_{\rm pl})=0$, while the other is  
to eliminate the threshold corrections to the $m_H$ from runnings of other couplings \cred{
to realize the nontrivial ultraviolet (UV) fixed points}.   
The former 
can be realized by a dynamical cancellation at $\Lambda_{\rm pl}$ 
over Planckian contribution, as has been argued in~\cite{Shaposhnikov:2009pv,Wetterich:2016uxm,Eichhorn:2017als,Pawlowski:2018ixd,Wetterich:2019qzx}. 
The latter is subject to some UV completion for the conformal SM 
embedded in a nontrivial-UV safety theory (called asymptotic safety), 
as has been extensively explored recently~\cite{Gies:2003dp,Shaposhnikov:2008xi,Gies:2009sv,Braun:2010tt,Bazzocchi:2011vr,Wetterich:2011aa,Antipin:2013pya,Gies:2013pma,Tavares:2013dga,Abel:2013mya,Litim:2014uca,Litim:2015iea,Bond:2016dvk,Pelaggi:2017abg,Bond:2017wut,Barducci:2018ysr,Eichhorn:2018yfc,Abel:2018fls}.  
With these dynamical conditions at hand, 
no corrections to the Higgs mass term can be generated at any loop order, 
which indeed gives rise to the quantum scale invariant SM at the infrared EW scale, 
thus, no fine-tuning or unnatural Higgs mass parameter will arise.

The key point here is to note that 
this kind of conformally extended SMs (embedded in the 
asymptotic safety) 
necessary includes one SM-singlet scalar, $S$, 
which couples to the Higgs doublet via biquadratic forms 
\textcolor{black}{with 
a real scalar~\cite{Hempfling:1996ht}
or an extra $U(1)$-charged scalar~\cite{Meissner:2006zh}, 
or a generic complex scalar with or without $CP$ violation~\cite{AlexanderNunneley:2010nw,Farzinnia:2013pga,Gabrielli:2013hma}, 
like $ |H|^2 S^2$}, known as the   
{\it Higgs portal scenario}. 
\cred{Then 
the renormalization group evolution of $\lambda_H$ necessarily receives
positive contributions from such a portal couplings and makes the $\lambda_H$ bounded from zero, to attain the stable EW vacuum.}  
In this case, the EW scale 
can be dynamically generated from 
the scale-symmetry breaking at the quantum level, 
in the following two different ways: 
\begin{itemize} 
\item 
{\it perturbative type} --   
Coleman-Weinberg (CW) mechanism~\cite{Coleman:1973jx,Gildener:1976ih}
for weakly-coupled massless scalars (\cite{Hempfling:1996ht,Meissner:2006zh,AlexanderNunneley:2010nw,Farzinnia:2013pga,Gabrielli:2013hma} and also 
see, e.g., related references that have cited those papers); 
\item 
{\it nonperturbative type} --  
dimensional transmutation of a nonperturbatively created scale 
by a strongly coupled hidden sector~%
\cite{Kubo:2014ova,Hur:2011sv,Holthausen:2013ota,Heikinheimo:2013fta,Hambye:2013sna,Kubo:2015cna,Kubo:2016kpb,Kubo:2017wbv,Ouyang:2018eub}.  
\end{itemize}
The scenarios of this class can therefore be called a 
{\it Higgs-portal scalegenesis}.

A common feature that all perturbative-type Higgs-portal scalegenesis mechanisms share 
is the presence of a flat direction in the tree-level scalar potential. 
It is necessarily present 
if the determinant of the scalar quartic-coupling matrix vanishes~\cite{Kannike:2019upf}. 
It ensures that 
all scalars acquire their VEVs simultaneously 
by quantum corrections, i.e. the CW mechanism, 
and thus, 
an inevitably light dilaton-like scalar state emerges 
as a pseudo-Nambu-Goldstone (NG) boson associated 
with the anomalous scale symmetry~\cite{Coleman:1973jx} 
(also called the scalon in the original Gildener-Weinberg approach~\cite{Gildener:1976ih}).

Even in the nonperturbative-type of the Higgs-portal scenario, 
similar flat directions can be observed in terms of   
low-energy effective scalar potential, 
where the Higgs portal interactions is established between the SM-like Higgs and 
composite scalars generated by the underlying strong dynamics. 
The scale symmetry breaking generated nonperturbatively looks 
built-in at the tree-level in the low-energy effective scalar theory, though.

Note that the Higgs portal scalegenesis posses universal experimental evidences: 
the predicted 
dilaton, arising as a singlet scalar fluctuation mode from the portal field 
$S$, will couple to SM particles 
due to the mixing with the SM-like Higgs boson $h$ through the Higgs portal 
interaction $|H|^2 |S|^2 = v_S v \, hs  + \cdots$ 
with the vacuum expectation values 
of the $S$ and $H$, $v_S$ and $v$. 
The size of the dilaton coupling to SM particles as well as 
the 125 GeV Higgs couplings to them are  
then universally controlled by the mixing angle 
$\theta$, respectively, 
in which the latter 
has severely been constrained by 
the Higgs coupling measurement experiments as $|\sin \theta| \lesssim 0.3$~\cite{Aad:2015pla}.  
When the dilaton mass is on the order of EW scale, or higher, 
it is mainly produced by gluon fusion process 
 and decays to the EW-dibosons $WW$ and $ZZ$ at the 
hadron colliders, like LHC. 
The possible excess events, which are 
$\propto \sin^2 \theta \times$ SM-like Higgs events    
at the invariant mass around the EW scale 
in those diboson channels, 
will thus be a generic prediction 
in the Higgs portal scenario.

In addition to the diboson signatures,  
the Higgs potential structure including the higher order terms in 
the $h$ field, such as the cubic $h^3$ term,  
will be modified from the SM prediction, which 
is parametrized by functions of $\theta$, 
with the ratio of $v_s/v$, 
and will be subject to the light dilaton resonance coupled to 
the diHiggs in the trilinear Higgs amplitude through the $h$-$s$ conversion process like $h^{(*)} \to s^{(*)} \to hh$. 
Note also that the dilaton resonance 
is generically narrow due to the small coupling strength 
to SM particles set by 
the phenomenologically small $\sin \theta$. 
Thus the diHiggs signatures will also be a characteristic 
probe for this scenario, 
as has been discussed by means of a specific Higgs portal scalegenesis~\cite{Ouyang:2018eub}.

Those signals can be predicted 
no matter what kind of ways is investigated to realize the EWSB via 
Higgs portal interactions  
with somewhat a light and narrow enough scalar  
(at some decoupling limit for heavier particles, if any), 
hence are universal predictions expected in an energy range within the reach of collider experiments~\cite{Ouyang:2018eub}. 
Though not having been analyzed, it is obvious that other models regarded 
as the Higgs portal scalegenesis~%
\cite{Coleman:1973jx,Gildener:1976ih,Hempfling:1996ht,Meissner:2006zh,AlexanderNunneley:2010nw,Farzinnia:2013pga,Gabrielli:2013hma,Kubo:2014ova,Hur:2011sv,Holthausen:2013ota,Heikinheimo:2013fta,Hambye:2013sna,Kannike:2019upf,Kubo:2015cna,Kubo:2016kpb,Kubo:2017wbv} can generically predict similar collider signatures.

Thus this kind of the conformal extension for the SM, 
namely, the Higgs-portal scalegenesis is thought to form  
a {\it universality class}, in a sense of the universal low-energy effective theory 
and related phenomenology.

Even in such a Higgs-portal scalegenesis, 
actually, the main focus on the realization of the EWSB is just going to be moved 
from the origin of the Higgs mass itself to the origin of the portal coupling 
because the latter somehow has to be ``negative''. 
Even working in CW mechanism~\cite{Coleman:1973jx}, 
one needs to require the portal coupling to be ``negative'' by hand,  
otherwise, none of the models can realize the EWSB (see, e.g., \cite{Sannino:2015wka}, 
and references therein)~\footnote{It has been discussed in~\cite{Iso:2012jn,Hashimoto:2014ela} 
that without a bare Higgs portal coupling, 
a mixing effect between the hypercharge gauge and a newly introduced 
gauge $(B-L)$ bosons can radiatively generate the portal coupling between 
the $B-L$ Higgs (regarded as a dilaton in that case) and the SM-like 
Higgs at the two-loop level. 
However, because of the higher loop-induced coupling, 
its size is quite small ($\sim O(10^{-3})$), 
which is required to realize the $B-L$ breaking at TeV scale, 
hence the mixing strength with the SM-like Higgs gets small enough as well, 
so that the predicted light dilaton couplings to 
diEW and diHiggs bosons will be negligibly smaller than 
other models having the sizable (negative) Higgs portal coupling 
(by hand) at tree-level. 
To this respect, we may exclude this kind of 
radiative generation scenarios from the universality class that 
we presently work on.}: 
CW mechanism cannot simply be applied to generate the EW scale 
since required parameters cannot be compatible to the observed values.

This fact implies the necessity for the dynamical origin 
for both of the scale and Higgs portal coupling generations including the negative sign 
to answer the origin of mass as a scenario completion. 
Furthermore, it should give a definite phenomenological consequence 
distinguishable in a sense of a unified category for the Higgs-portal scalegenesis.

In this write-up, we demonstrate a universal interpretation of models 
leading to the Higgs-portal scalegenesis as a low-energy effective theory, 
which arises along a conformal/flat direction,  
with the EWSB structure similar to the SM encoded. 
This builds the universality class in the (nearly) conformal/flat direction 
including the SM, without loss of generality as will be seen below.  
We then present a discriminator for the universality class, which is to be closely related to the very origin of the negative Higgs mass term/origin of mass: 
{\it A dynamical origin of Higgs portal}.  


\section{A dynamical origin of Higgs portal: 
a generic low-energy description} 

To begin with, we show that 
a conventional scale-invariant Higgs portal scenario emerges in a decoupling limit 
for a scale-invariant realization of 
two Higgs doublet model with a light dilaton introduced. 
In addition, we see that in such a class of models, 
a softly broken $Z_2$/$U(1)_A$ for the Higgs sector 
plays the crucial role to realize the negative Higgs-portal coupling  
between the SM-like Higgs and the light dilaton.

Having in one's mind 
a scale-invariant realization of 
two Higgs doublet model with a light dilaton $(\chi)$, 
one finds the potential terms like  
\begin{align} 
V \ni \chi^2 \left[ c_0 |H_1|^2 + c_1 (H_1^\dag H_2 + {\rm H.c.}) + c_2 |H_2|^2 \right]
\,, 
\end{align}
where $c_{0,1,2}$ are arbitrary dimensionless coefficients, 
and $H_{1,2}$ are the Higgs doublets. 
To manifestly see a symmetry structure of current concern, 
one may introduce a two-by-two Higgs matrix form, $\Sigma = (H_1, H_2^c)$ 
(with $H_2^c$ being the charge conjugated field of $H_2$), 
to rewrite the terms as 
\begin{align} 
 V  \ni & \chi^2 
 \Bigg[ \left( \frac{c_0 + c_2}{2}\right) {\rm tr} \left[ \Sigma^\dag \Sigma \right]  
\notag \\ 
 & 
 + c_1 ({\rm det} \Sigma + {\rm H.c.}) 
 + \left( \frac{c_0 - c_2}{2} \right) {\rm tr} \left[ \Sigma^\dag \Sigma \sigma^3 \right] 
 \Bigg]
\,, 
\end{align}
where $\sigma^3$ is the third Pauli matrix. 
It is easy to see that the potential is built upon 
a global chiral $U(2)_L \times U(2)_R$ symmetry for the two Higgs flavors, 
where the $SU(2)_R$ part is in part explicitly broken down 
(to the subgroup corresponding to the third component of $SU(2)$) 
by the third term and the $U(1)_A$ part 
(which is usually called soft-$Z_2$ breaking term in the context of two-Higgs doublet models) 
is broken by the second $c_1$ term. 
The same chiral-two Higgs sector structure (without the scale invariance) 
has been discussed in~\cite{Hill:2019ldq,Hill:2019cce}.

At this point, the dimensionless couplings $c_{0,1,2}$  
are simply assumed to be real and 
positive to have a conformal/flat direction.  
In that case the conformal/flat direction for both the scale and EW breaking VEVs 
can be achieved, where  
the direction for the EW scale is somewhat deformed due to the 
mass mixing by $c_1$ as 
\begin{align} 
\tilde{v}_2 \equiv v_2 + (c_1/c_2) v_1 =0
\,. 
\end{align} 
Note that this deformation is nothing but a base transformation: 
$v_{1,2} \to \tilde{v}_1(=v_1), \tilde{v}_2$, 
hence can generally and smoothly be connected to the 
SM limit with the $v_1$ only.

Now, assume the maximal isospin breaking for the two Higgs doublets, 
where $c_0/c_2 \to 0$, and the soft enough $U(1)_A$/$Z_2$ breaking, 
by taking $c_1/c_2 \ll 1$. 
Then, one may integrate out the heavy Higgs doublet $H_2$ 
to get the solution for the equation of motion, $H_2 \approx - (c_1/c_2) H_1$~\footnote{
The $H_2$  mass term takes a $\chi$ field-dependent form like  $m_{H_2}^2(\chi) = c_2 \chi^2$, with $c_2 >0$ and ${\cal O}(1)$ as assumed in the text. This $m_{}H_2(\chi)$ becomes the $H_2$ mass after the $\chi$ develops the VEV $\langle \chi \rangle$, which is by construction lager than the EW scale, or the lightest Higgs mass identified as the 125 GeV Higgs's. 
Thereby, one can safely integrate out the heavy $H_2$ by taking into account its a priori heaviness.}. 
Plugging this solution back to the potential, one finds 
\begin{align} 
V \approx - \left(\frac{c_1^2}{c_2} \right) \chi^2 |H_1|^2 
\,, 
\end{align}  
which is nothing but a desired Higgs portal model, where the portal coupling 
$\lambda_{H \chi} = - c_1^2/c_2$ has been dynamically induced 
{\it including the minus sign} without assuming anything, 
which is reflected by the attractive interaction of the scalar-exchange induced potential 
in a sense of the quantum mechanics.  
One should also realize that the small portal coupling can actually be 
rephrased by the small size of the soft-$Z_2$/$U(1)_A$ breaking for the underlying 
two Higgs doublet model. 
Note also that the conformal/flat direction oriented in the original two Higgs doublet model  
is smoothly reduced back to the one in the Higgs portal model, as it should be.

This generation mechanism is nothing but what is called the bosonic seesaw~%
\cite{Calmet:2002rf,Kim:2005qb,Haba:2005jq,Antipin:2014qva,Haba:2015lka,Haba:2015qbz,Ishida:2016ogu,Ishida:2016fbp,Haba:2017wwn,Haba:2017quk,Ishida:2017ehu}, 
which one can readily check if 
the scalar mass matrix takes the seesaw form, 
namely, its determinant is negative 
under the assumption made as above. 
Note also that the original conformal/flat direction $\tilde{v}_2 \equiv v_2 + (c_1/c_2) v_1 =0$ can  
actually be rephrased in terms of the bosonic seesaw relation: 
when the mixing is small enough (i.e. $c_1/c_2 \ll 1$), 
the heavy Higgs partner via the bosonic seesaw approximately arises 
as $\tilde{H}_2 \simeq H_2 + (c_1/c_2) H_1$, 
so the conformal/flat direction has been realized 
due to the presence of an approximate inert $H_2$. 
Thus, the bosonic seesaw provides the essential source for the Higgs-portal scalegenesis, 
to predict the universal low-energy new-physics signatures like 
significant deviations for Higgs cubic coupling measurements compared with 
the SM prediction, and for the light dilaton signatures in diHiggs, diEW bosons, 
as aforementioned above.

\section{A very origin of Higgs portal: a UV completion}

One can further observe that 
a hidden strong gauge dynamics -- often called hidden QCD  
(hQCD) or hypercolor~\cite{Haba:2015qbz,Ishida:2016ogu,Ishida:2016fbp,Ishida:2017ehu} -- 
provides the dynamical origin for the softly-broken $Z_2$ or $U(1)_A$ symmetry 
and an alignment to the flat direction, 
which are just given as ad hoc assumptions 
in the framework of the scale-invariant realization of 
two Higgs doublet model, as done right above.  
Indeed, a class of the hQCD as explored in~\cite{Haba:2015qbz,Ishida:2016ogu,Ishida:2016fbp,Ishida:2017ehu} 
can dynamically generate a composite dilaton (arising generically as an admixture of 
fluctuating modes for hQCD fermion bilinear, 
like a conventional sigma meson in QCD, and gluon condensates such as a glueball~\footnote{Even in a naive scale-up version of QCD with the small number of flavors 
as applied in the literature~\cite{Ishida:2016ogu,Ishida:2016fbp,Ishida:2017ehu}, 
it has recently been argued~\cite{Alexandru:2019gdm} 
that there might exist an infrared conformality, 
supporting the QCD dilaton to be light enough, compared to the 
dynamical intrinsic scale. 
Even if it is not the case, 
the hQCD flavor structure can straightforwardly be extended 
from the three flavor to many flavors, say, eight's~\cite{Aoki:2014oha,Aoki:2016wnc}, 
with keeping the bosonic seesaw mechanism, 
so that a manifest light composite dilaton can be generated 
by the nearly conformal dynamics, as has recently been discussed~\cite{Ishida:2019wkd}}.).

Consider an hQCD with three colors and three flavors, as a minimal model 
to realize the bosonic seesaw as discussed in~\cite{Haba:2015qbz,Ishida:2016ogu,Ishida:2016fbp}, 
where the hQCD fermions form the $SU(3)$-flavor triplets, 
$ F_{L,R} = (\Psi_i, \psi)^T_{L,R}$, 
having the vectorlike charges with respect to the SM gauges like, 
$\Psi_{i(i=1,2)} \sim (N, 1, 2, 1/2)$, and $\psi \sim (N, 1, 1, 0)$, 
for the hQCD color group, $SU(N=3)$, and $SU(3)_c \times SU(2)_W \times U(1)_Y$. 
Thus this hQCD possesses the (approximate) 
``chiral" $U(3)_{F_L} \times U(3)_{F_R}$ symmetry as well as the classical-scale invariance, 
the former of which is explicitly broken by the vectorlike SM gauges. 
Besides, we shall introduce the following terms, which are SM gauge-invariant 
but explicitly break the chiral symmetry: 
$
{\cal L}_{y_H}  
= 
- 
y_H 
\, 
\bar{F}_L \cdot   
\left( 
\begin{array}{cc} 
0_{2\times 2} & H \\ 
H^\dag & 0   
\end{array} 
\right) 
\cdot 
F_R 
+ {\rm H.c.} 
$. 
Note that  in addition to this $y_H$-Yukawa term, 
the $U(1)_{F_A}$ symmetry is explicitly broken also by the anomaly coupled to hQCD gluons, 
which can, however, be transferred to this $y_H$-Yukawa term 
by the $U(1)_{F_A}$ rotation, 
so that it fully controls the size of the $U(1)_{F_A}$ symmetry breaking.

The remaining (approximate) chiral 
$SU(3)_{F_L} \times SU(3)_{F_R} (\times U(1)_{F_V})$ symmetry 
is broken by the chiral condensate, invariant under the SM gauge symmetry, 
$
\langle \bar{F}F  \rangle 
= 
\langle \bar{\Psi}_i \Psi_i \rangle 
= 
\langle \bar{\psi} \psi \rangle \neq 0
$, 
down to the diagonal subgroup $SU(3)_{F_V} (\times U(1)_{F_V})$ 
at the scale $\Lambda_{\rm hQCD}$, 
just like the ordinary QCD. 
This spontaneous chiral breaking thus leaves us as the low-energy spectrum 
with the eight NG bosons.

The low-energy description for the ${\cal L}_{y_H}$,  
below the scale  $\Lambda_{\rm hQCD}$, can look like 
\begin{align} 
& 
\chi^2 \left[ c_1 (H_1^\dag \Theta + {\rm H.c.}) + c_2 |\Theta|^2    \right]
\notag \\ 
= & \chi^2 
\left\{
 c_1 ({\rm det} \Sigma + {\rm H.c.}) 
 + c_2  {\rm tr}\left[\Sigma^\dag \Sigma \left(\frac{1-\sigma^3}{2} \right)\right] 
 \right\}
\,, 
\end{align}
where $\Sigma = (H, \Theta^c)$ with $\Theta \sim \bar{\psi}_R \Psi_L$ 
being a composite Higgs doublet~\footnote{
When one works on 
hQCD theory with hQCD fermions in higher dimensional representations, like a real or a pseudo-real representation, 
the seesaw partner $\Theta$ would be a composite Nambu-Goldstone Higgs-doublet,
as employed in~\cite{Haba:2017quk}. 
In that case,  one would have $c_2 =0$ at the $\Lambda_{\rm hQCD}$ scale. 
Going down to lower scales, however, EW radiative corrections would generate 
the $\Theta$ mass on the order of ${\cal O}(g_W/(4\pi) \Lambda_{\rm hQCD})$, where $\Lambda_{\rm hQCD} = {\cal O}(1)$ TeV, as will be seen from the phenomenological bounds later. 
Hence, this $\Theta$ mass scale should be of ${\cal O}(100)$ GeV, less than 
the EW scale and smaller than a composite dilaton $(\chi)$ mass (predicted at around 300 GeV~\cite{Ouyang:2018eub}). 
Therefore, one cannot integrate out 
the $\Theta$, instead, the dilaton $\chi$ will be integrated out 
such that 
the theory will be away from 
the conformal direction. In other words, this hQCD model does not blelong to 
the universality class in which the Higgs portal between the SM-like Higgs 
$H_1$ and a SM-singlet dilaton $\chi$ 
is necessarily present at the low-energy theory. 
This is the case for a minimal setup only with the HC theory and the $y_H$-like Yukawa term as well as the SM gauge interactions.   
Going beyond the minimal setup could make the theory come back on the track of the conformal direction.
}; 
$c_1 \simeq y_H$ up to some renormalization effect scaling down to the scale $\Lambda_{\rm hQCD}$; 
$c_2$ has been generated by the chiral condensate 
$\langle \bar{F}F  \rangle$ scaled by the VEV of the composite hQCD dilaton $\chi$. 
This is nothing but the form for a scale-invariant two Higgs doublet model as discussed above, 
so the bosonic seesaw should work, 
to bring the theory back to the Higgs portal model as the low-energy description. 
Of importance is to note also that 
the approximate inertness of the second Higgs doublet, 
necessary to have the conformal/flat direction, 
is now manifest because of the 
robust Vafa-Witten theorem~\cite{Vafa:1983tf}, 
which ensures the zero VEV for the non-vectorlike condensate, 
such as $\Theta \sim \bar{\psi}_R \Psi_L$, in this vectorlike hQCD, 
and the positiveness of the $c_2$ 
(i.e. the positive mass square of the $\Theta$) 
as long as the chiral manifold describing the low-energy hQCD is stable.

\section{Discriminating the universality class}

One may identify the $U(1)_{F_A}$ in the hQCD 
as the $U(1)_A$ for the previous two Higgs sector. 
Then one can say that the ad hoc assumption 
(the soft-$Z_2$/$U(1)_A$ breaking by taking 
$c_1/c_2 \ll 1$ and maximal isospin breaking for the Higgs sector: $c_0=0$) 
is naturally realized by the hQCD, in which the bosonic seesaw mechanism is built, 
where the smallness of the $c_1$ can be understood by 
the existence of light hQCD pions~\footnote{Although the $y_H$ gives a tachyonic mass to the lightest neutral hQCD pion, 
one can resolve it by introducing extra singlet pseudoscalar as discussed 
in~\cite{Ishida:2016ogu,Ishida:2016fbp,Ishida:2017ehu} 
without conflicting any discussions in the present paper.}.  
Thus, the origin of the EWSB derived from the negative portal coupling 
is tied to the explicit-hidden chiral symmetry-breaking (and/or $U(1)_A$ breaking) 
in the hQCD sector.

The small $y_H$ coupling can lead to the custodial symmetry breaking, 
the oblique correction such as the $T$-parameter constraint has to be discussed 
due to corrections from EW-charged hQCD pions. 
Such EW charged pions also significantly contribute to the 
125 GeV Higgs decay to diphoton, in addition to the 
overall suppression by the mixing angle with the light dilaton, which is universally 
present in the Higgs-portal scalegenesis.     
We have confirmed sufficiently allowed parameter spaces under those constraints, 
which will be in detail reported in another publication. 
For instance, when we take $\Lambda_{\rm hQCD}=1(2)$ TeV, 
the EW-charged hQCD pion mass is bound to be $\gtrsim 450(700)$ GeV 
for the Higgs-dilaton mixing strength $\sin^2\theta=0.1$, and 
$\gtrsim 400(600)$ GeV for $\sin^2\theta=0.05$, 
along with the soft-$Z_2/U(1)_A$ breaking coupling $y_H \lesssim 0.1$, 
which yields the Higgs portal coupling $\lambda_{H\chi} \lesssim 0.1$,  
and the hQCD dilaton having the mass around 300 GeV as in~\cite{Ouyang:2018eub}.     
Such light pions can be produced at the LHC by EW interactions (vector boson fusions) 
via the chiral anomaly in the hQCD, so that the predicted production cross sections 
will be quite small (roughly at most $\sim 10^{-1}$ fb at 13 TeV) due to 
the loop factor suppression, which is compared with the currently stringent 
upper bound $\sim 10^2$ fb at the corresponding mass range~\cite{Sirunyan:2019der}, 
and  
it may be hard to directly detect even in the high-luminosity epoch 
(for similar EW-charged pion signals, e.g., see~\cite{Matsuzaki:2016joz}). 
Note even in that case that 
the presently proposed hQCD can be probed by correlated deviations for 
the 125 GeV Higgs decay to diphoton by hQCD pion loops and the diboson channels 
including diHiggs and di-EW bosons, as discussed in~\cite{Ouyang:2018eub}, 
which are definitely characteristic in the universality class of the Higgs-portal scalegenesis.

Thus, the light hQCD pions will be the definite discriminator 
for the universality class of the Higgs portal scalegenesis.  
If both a light dilaton and hQCD pions 
(whose masses are expected to be around/below TeV scale) 
are detected at forthcoming collider experiments, 
it would be the hQCD that gives the very origin of the Higgs portal coupling, 
hence the very origin of Higgs mass term.   
In addition, $y_H$ term which breaks chiral symmetry explicitly potentially induces 
a significant deviation of the $T$ parameter~\cite{Peskin:1990zt,Peskin:1991sw}. 
Thus, the EW precision data also provide some hints to explore the models of this 
universality class.

\begin{figure}
\begin{center}
\includegraphics[width=7cm,clip]{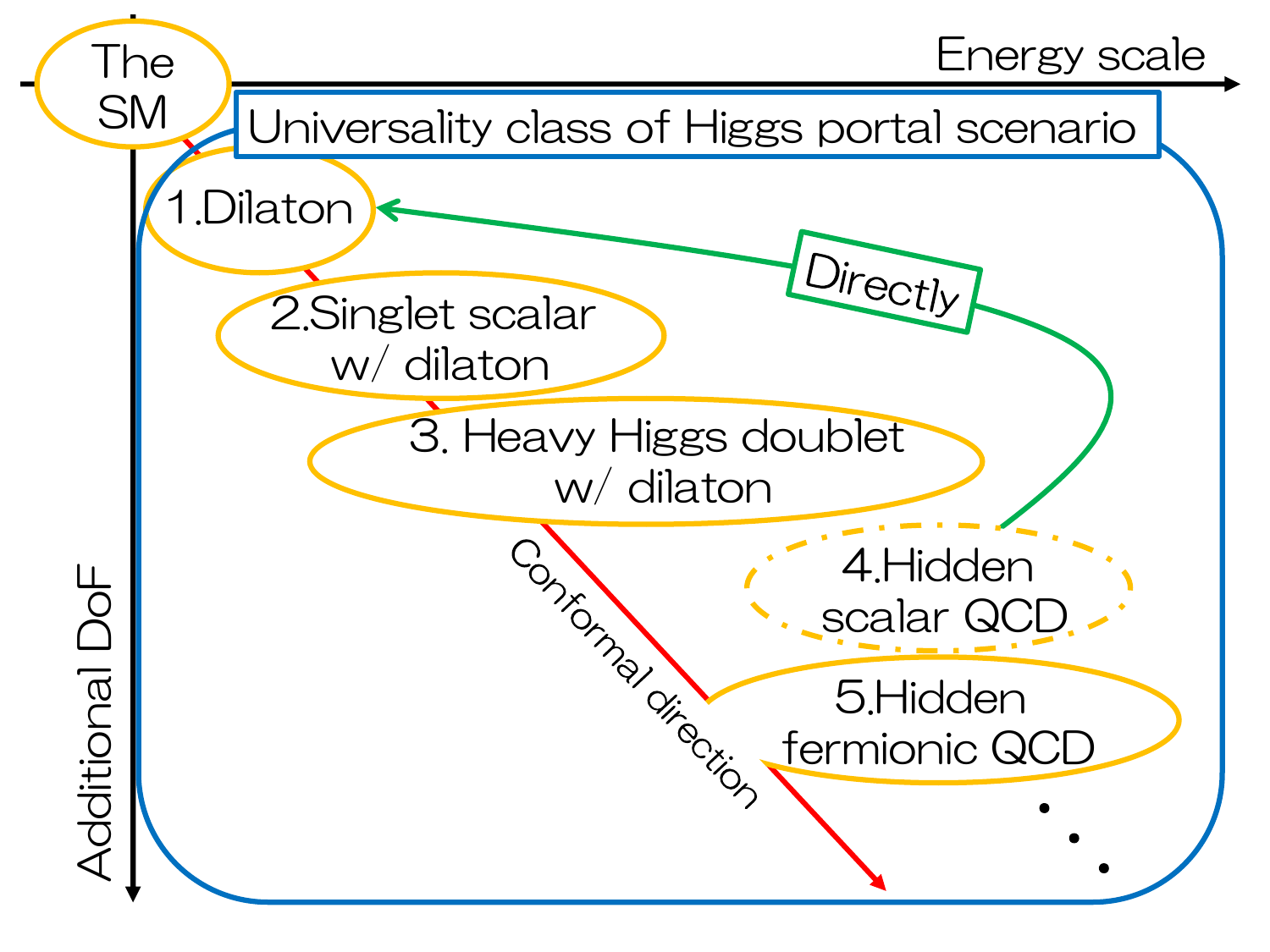}
\end{center}
\caption{
A schematic picture of the universality class for Higgs portal scenarios. 
All the extensions correspond to 1).~\cite{Coleman:1973jx,Gildener:1976ih}, 
2)~\cite{Hempfling:1996ht,Meissner:2006zh,AlexanderNunneley:2010nw,Farzinnia:2013pga,Gabrielli:2013hma}, 
3).~\cite{Haba:2015lka,Hill:2019ldq,Hill:2019cce}, 
4).~\cite{Kubo:2014ova,Holthausen:2013ota, Kubo:2015cna,Kubo:2016kpb,Kubo:2017wbv,Ouyang:2018eub}, 
and 
5).~\cite{Haba:2005jq,Hur:2011sv,Antipin:2014qva,Haba:2015qbz,Ishida:2016ogu,Ishida:2016fbp,Haba:2017wwn,Haba:2017quk,Ishida:2017ehu}, 
respectively. 
All categorized models predict the same low-energy phenomenology 
related to the Higgs physics, like those noted in the main text. 
A dynamical origin for this universality class can be encoded in a sort of fermionic hQCD (\#5 in the figure)
with distinct light pion signatures as well as the universal Higgs-related ones.}
\label{universal}
\end{figure}


\section{Conclusion} 

In conclusion, 
the universality class that is presently proposed can be depicted as in Fig.\ref{universal}. 
The universality class and its disentangled  
dynamical origin would give a novel guideline 
along the conformal extension of the SM leading 
to a possible solution of the longstanding quest on the gauge hierarchy (fine tuning) problem. 
This would also give a clear understanding of the hidden new physics 
in searching for the dynamical origin of the Higgs sector, hence the origin of mass, 
which are to be testable in upcoming collider experiments.

More detailed studies regarding distinct collider signatures 
for the two-Higgs doublet model-type and hQCD type are worth performing, 
to be pursued elsewhere. 
Also, the thermal histories as well as possible gravitational wave signals 
for this universality class 
could be discriminated, which would deserve to the future research direction.


\section*{Acknowledgements} 
 
We are grateful for Kristjan Kannike 
and Jiang-Hao Yu for fruitful discussions. 
This work was supported in part by the National Science Foundation of China (NSFC) under Grant No. 11747308, 
and the Seeds Funding of Jilin University (S.M.).  
H.I. and R.O. thank for the hospitality of Center for Theoretical Physics and College of Physics, 
Jilin University where the present work has been partially done. 
The work of H.I. was partially supported by JSPS KAKENHI Grant Numbers 18H03708. 
R.I. was supported by the Estonian Research Council grant PRG434. 



\end{document}